\documentclass[reprint,superscriptaddress,amsmath,amssymb,prl,nolongbibliography]{revtex4-2}

\usepackage{graphicx}
\usepackage{xcolor}
\usepackage{dcolumn}
\usepackage{bm}

\newcommand{\me}{m_{\rm e}}
\newcommand{\mi}{m_{\rm i}}
\newcommand{\nue}{\nu_{\rm e}}
\newcommand{\nui}{\nu_{\rm i}}
\newcommand{\fo}{f^{(0)}}
\newcommand{\fl}{f^{(1)}}
\newcommand{\wi}{\omega_{\rm pi}}
\newcommand{\we}{\omega_{\rm pe}}
\newcommand{\Rw}{\mathcal{R}_{\rm w}}
\newcommand{\half}{\frac{1}{2}}

\begin{document}

\preprint{APS/123-QED}

%%%%%%%%%%%%%%%%%%%%%
% Title
%%%%%%%%%%%%%%%%%%%%%
\title{Electron heating in high Mach number collisionless shocks}

%%%%%%%%%%%%%%%%%%%%%
% Authors
%%%%%%%%%%%%%%%%%%%%%
\author{A. Vanthieghem}
\email{arno.vanthieghem@obspm.fr}
\affiliation{Sorbonne Universit\'e, Observatoire de Paris, Universit\'e PSL, CNRS, LERMA, F-75005, Paris, France}
\affiliation{Department of Astrophysical Sciences, Princeton University, Princeton, NJ 08544, USA}
\affiliation{Department of Astro-fusion Plasma Physics (AFP), Headquarters for Co-Creation Strategy, National Institute of Natural Sciences, Tokyo 105-0001, Japan}
 
\author{V. Tsiolis}
\affiliation{Department of Astrophysical Sciences, Princeton University, Princeton, NJ 08544, USA}

\author{A. Spitkovsky}
\affiliation{Department of Astrophysical Sciences, Princeton University, Princeton, NJ 08544, USA}

\author{Y. Todo}
\affiliation{National Institute for Fusion Science, National Institutes of Natural Sciences, Toki, Gifu 509-5292, Japan}

\author{K. Sekiguchi}
\affiliation{Department of Astro-fusion Plasma Physics (AFP), Headquarters for Co-Creation Strategy, National Institute of Natural Sciences, Tokyo 105-0001, Japan}

\author{F. Fiuza}
\affiliation{GoLP/Instituto de Plasmas e Fusão Nuclear, Instituto Superior Técnico, Universidade de Lisboa, 1049-001 Lisbon, Portugal}
\affiliation{High Energy Density Science Division, SLAC National Accelerator Laboratory, Menlo Park, California 94025, USA}

\date{\today}

%%%%%%%%%%%%%%%%%%%%%
% Abstract
%%%%%%%%%%%%%%%%%%%%%
\begin{abstract}
The energy partition in high Mach number collisionless shock waves is central to a wide 
range of high-energy astrophysical environments. We present a new theoretical model for electron 
heating that accounts for the energy exchange between electrons and ions at the shock. The 
fundamental mechanism relies on the difference in inertia between electrons and ions, resulting 
in differential scattering of the particles off a decelerating magnetically-dominated microturbulence 
across the shock transition. 
We show that the self-consistent interplay between the resulting ambipolar-type electric field and diffusive transport of electrons 
leads to efficient heating in the magnetic field produced by the Weibel instability in the high-Mach number regime and is consistent with fully kinetic simulations.
\end{abstract}

\maketitle

%%%%%%%%%%%%%%%%%%%%%
% Introduction
%%%%%%%%%%%%%%%%%%%%%
High Mach number collisionless shocks shape the electromagnetic signatures of many astrophysical 
environments. From pc-scale young supernovae remnants to Mpc-scale virial rings of galaxy clusters, 
the emission relies on the efficient acceleration of electrons and ions to highly relativistic speeds 
at the interface between a supersonic flow and a weakly magnetized plasma. The injection of electrons 
into acceleration processes~\cite{Krimskii_1977, Bell_1978_I, Bell_1978_II, Blandford_1978, Drury_1983, Blandford_1987} 
and the interpretation of observations~\cite{Nikolic_2013, Miceli_2019, Reynolds_2021} 
directly depend on the electron heating efficiency and properties. The mechanisms that underpin the energy 
transfer and the temperature ratio between electrons and ions thus constitute one of the most fundamental 
open questions in our understanding of these blast waves.  

Left as a free parameter from the Rankine-Hugoniot jump conditions, the electron-to-ion temperature ratio 
is inferred via various observational probes~\cite{Ghavamian_2013, Raymond_2023},  from radio and X-ray 
synchrotron emissions within young Supernova Remnant (SNR) shocks ($M_A\,\gtrsim\,10^2$)~\cite{Reynolds_2008} 
to in-situ measurements at Earth's bow shock ($M_A\,\lesssim\,10$)~\cite{Feldman_1982}. 
The latter allowed for direct characterization of the shock dynamics with essential results on the structure 
gleaned from the Magnetospheric Multi-Scale (MMS) spacecraft~\cite{Johlander_2023}, but the direct 
characterization of the temperature ratio for $M_A\,\gtrsim\,10^2$ remains elusive.
In parallel, rapid developments in high-power lasers, such as at OMEGA and the National Ignition Facility, 
are opening valuable opportunities to investigate high-Mach number collisionless shocks in controlled 
laboratory experiments for direct measurement of particle energization 
processes~\cite{Huntington_2015, Schaeffer_2017, Fiuza_2020, Grassi_2021}. 

These observational and experimental studies are supported by significant joint numerical efforts to 
self-consistently model the shock dynamics through multi-dimensional kinetic simulations, 
providing a detailed characterization of the plasma processes at play over limited timescales.  
These simulations cover various regimes of magnetosonic Mach 
numbers~\cite{Kato_2008, Amano_2009, Kato_2010, Kumar_2015, Guo_2017, Guo_2018, Kang_2019, Tran_2020, Bohdan_2021, Lezhnin_2021, Morris_2023}. 
Despite all these substantial efforts, identifying and modeling the dominant source of electron 
heating in high magnetosonic Mach number shocks remains a critical challenge.

%%%%%%%%%%%%%%%%%%%%%
% Content
%%%%%%%%%%%%%%%%%%%%%
In this Letter, we present a model for electron transport and heating in self-generated microturbulence 
that can accurately capture the electron-ion temperature ratio observed in fully kinetic simulations 
of high-Mach number shocks. The generality of the approach presented here opens new avenues 
for modeling energy partition in systems governed by magnetically dominated turbulence. 

\begin{figure}
  \centering
  \includegraphics[width=1\columnwidth]{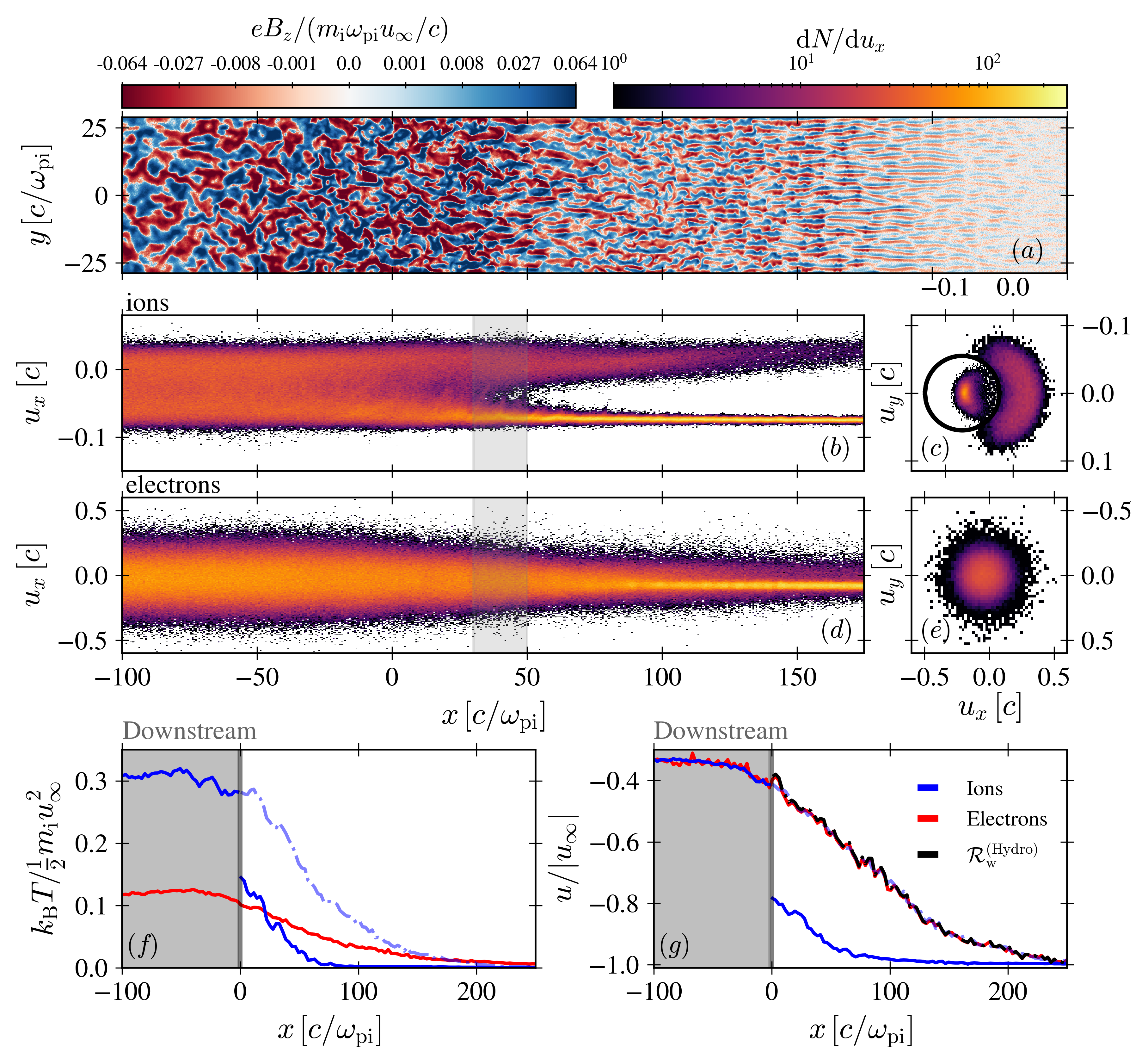}
  \caption{The characteristic structure of an unmagnetized electron-ion collisionless shock wave with
  $u_\infty\,=\,-0.075$, $k_{\rm B} T_\infty\,=\,6.8\,\times\, 10^{-4} \ \times \frac{1}{2}m_i u_\infty^2$
  ($M\simeq40$)
  and mass ratio $\mi/\me\,=\,49$, at time $\wi t \approx 6.1\times10^3$. The turbulent magnetic field is shown in (a), the $x-u_x$ phase space profile is shown in (b) for the ions, and (d) for the electrons. Insets (c) and (e) show the respective momentum distribution corresponding to the shaded area of panels (b) and (d). The circle in (c) differentiates the backstreaming beam and background ions. The temperature profile, shown in (f), shows the characteristic heating of the background ions (solid blue) and electrons (solid red) up to the downstream (shaded). For comparison, the total ion temperature corresponds to the dot-dashed line. The velocity profile for each species is shown in (g). The black line 
  corresponds to the numerical estimate of the turbulence frame.}
  \label{fig:fig1}
\end{figure}

We start by analyzing the typical structure of high-Mach number collisionless shocks 
from the results of large-scale fully-kinetic plasma simulations, which can self-consistently 
capture the dynamics of shock formation and particle heating. We have performed a 
series of 2D particle-in-cell (PIC) simulations with the relativistic electromagnetic 
code TRISTAN-MP~\cite{Anatoly_2005}. We initialize an electron-ion plasma at rest 
and set the left reflecting and conducting wall of our simulation box in motion 
along $+\hat{\mathbf{x}}$. The interaction between the reflected beam and the 
stationary plasma results in the formation of a shock. Our simulations are performed 
in the upstream frame. All quantities are then transformed and presented in the 
shock-front frame. A moving injector gradually recedes from the left wall at the 
speed of light.  Space and time coordinates are normalized to the ion plasma skin 
depth $c/\omega_{\rm pi}$ and frequency 
$\omega_{\rm pi} = \sqrt{4\pi n_\infty e^2/m_{\rm i}}$, with $n_\infty$ the proper 
upstream density. We adopt a resolution of $10$ cells per electron skin depth, 
$\Delta x =  \Delta y = 0.1\, c/\omega_{pe}$, and the timestep is 
$c\Delta t$ = $0.45 \, \Delta x$. In each direction, we use 32 particles per cell (ppc) 
and filter the current 32 times per timestep~\cite{Anatoly_2005}.  
We tested convergence varying ppc $=\,[32,128]$, $\frac{\we}{c}\Delta x\,=\,[0.05,0.1]$, 
and $|u_\infty|\,=\,[0.075, 0.2]c$, where $u_\infty$  is the far upstream velocity 
in the shock front frame (see Supplemental Material~\cite{Supp_Mat}).

As a reference for the following discussion, we consider a non-relativistic 
shock in the limit of very high Alfvén Mach number, i.e., initially 
unmagnetized, with $u_{\infty} = -0.075 c$, $m_i/m_e = 49$, and upstream ion 
temperature $k_B T_{i, \infty} = 1.91\!\times\! 10^{-6}\, m_i c^2$. The results 
of the simulation are illustrated in Fig. \ref{fig:fig1}, where we zoom in on 
the shock structure and precursor region. We observe that the interaction 
between the shock-reflected hot beam of ions propagating at positive velocity 
and the incoming upstream plasma drives a microturbulence 
via the Weibel, or current-filamentation, 
instability~\cite{Weibel_1959,Fried_1959} (Fig.\ref{fig:fig1}a) and 
leads to efficient electron heating (Fig.\ref{fig:fig1}f-g) to an 
electron-to-ion temperature ratio of $T_{\rm e}/T_{\rm i}\,\sim\,0.3$. 
Qualitatively, our model captures the deceleration of the 
turbulence across the shock and the associated charge separation between 
species of different inertia.
Electrons accelerate in the coherent electrostatic field that ensues and 
isotropize over their short scattering time scale through fast decoherence 
of the betatron motion. This diffusive process leads to efficient energy 
channeling between electrons and ions.

The microturbulence is magnetically dominated, so the scalar 
$\bm{E}^2 - \bm{B}^2 < 0$ everywhere in the shock precursor and 
downstream. This means one can always find a frame, $\Rw$, in which 
the electric field component vanishes locally. This frame drifts at 
a velocity $u_{\rm w}$ in the shock front frame. For statistically 
homogeneous turbulence transverse to the shock normal, the instantaneous 
velocity of this frame, $u_{\rm w}$, is a function of the longitudinal 
$x$-coordinate only.  The scattering center frame $\Rw$ extracted from 
the fully hydrodynamic limit~\cite{Pelletier_2019} shows
good agreement of the average proper motion of nonlinear structures, 
close to the electron drift velocity (Fig.\ref{fig:fig1}g). Ions from the backstreaming beam 
can be differentiated from the background --- i.e., incoming upstream flow 
--- via a threshold set at 
$|u - u_\infty|^2\,\lesssim\,u_{\rm thr}^2\,\simeq\,\frac{1}{2}\,u_\infty^2$ (Fig.\ref{fig:fig1}c). 
Across the shock transition, $\Rw$ does not coincide with the drift velocity of the background 
ions (Fig.\ref{fig:fig1}g) and is, therefore, non-ideal. Motivated by 
these observations, we explore a theoretical description of electron 
transport and heating in high-Mach number shocks to model the fraction 
of incoming energy density $\half n_\infty\mi u_\infty^2$ imparted to 
the electron distribution.

The equation of motion for a single charged particle in a non-inertial frame reads 
$\dot{\bm{p}} \,=\, \bm{p} \cdot \, \delta\hat{\Omega}_t + q\, \bm{E} - m \dot{\bm{u}}_{\rm w}$, 
where the first term accounts for pitch-angle variation in $\Rw$, the second term 
accounts for acceleration by the longitudinal electric field $\bm{E}$, and the 
last term for the non-inertial nature of $\Rw$. This approach offers a natural way to disentangle the contribution of the motional and electrostatic electric fields. If one defines $\bm{p} \cdot \delta\hat{\Omega}_t$ 
as a random force with the stochastic variables encoded in the form of 
the rotation matrix $\delta\hat{\Omega}_t$ and keeps a self-consistent 
$\bm{E}$-field contribution, the above equation of motion reduces 
to a semi-dynamical model of transport in a Langevin equation~\cite{Balescu_1997}. 

At this stage, the electric field contribution and origin are still unclear. 
Following the same arguments as in~\cite{Lemoine_2019_PRL, Pelletier_2019, Vanthieghem_2022}, 
we assume that the variation timescale of the turbulent structures is much larger 
than the typical scattering time of the electrons off these quasi-magnetostatic structures. 
Therefore, we neglect the contribution of the inductive electric field from the linear 
growth of plasma instabilities. We constrain the electric field to be electrostatic 
and build a reduced description for electron heating along those lines. 
The broadband longitudinal and transverse spectra of Weibel modes excited 
in the shock precursor, coupled with the nontrivial contributions of other 
channels such as the definition of $\Rw$ and $\bm{E}$, make the statistical 
description of $\delta\hat{\Omega}_t$ challenging to compare with fully kinetic 
simulations. The simplest nontrivial approximation assumes an isotropic 
Gaussian white noise process where $\delta\hat{\Omega}_t$ is a linear 
combination of the generators of the rotation group corresponding to 
pitch-angle scattering in the turbulent magnetic field~\cite{Plotnikov_2011}. 
Based on these assumptions, we now aim at deriving a self-consistent
relation for the electrostatic coupling between electrons and ions.

In this limit, the stochastic differential equation of motion is equivalent to a transport 
equation for the particle distribution $f_s$ for the species $s$. In the shock-front 
frame, where the system is assumed stationary, the transport equation reads:
\begin{align}\label{eq:tr_eq}
   \left( m u_{\rm w}  + p^x \right) \partial_x f_s &-\,m\,\partial_x u_{\rm w} \left( m u_{\rm w} 
    + p^x \right)  \partial_{p^x} f_s \nonumber \\
   +\,m\,q E_x\,\partial_{p^x} f_s  
   &\,=\, \frac{m}{2} \partial_\mu \left[ \nu_s \left( 1 - \mu^2 \right) \right] \partial_\mu f_s \,,
\end{align}
where the right-hand side is the operator for elastic scattering in $\mathcal{R}_{\rm w}$
in 3D with $\mu\,=\,\cos\theta$ the pitch-angle cosine and $\nu_s$ 
is the scattering frequency~\cite{Webb_1989}. For a two-dimensional distribution, 
the operator reduces to $\partial_\theta\nu\partial_\theta f_s$.
In the nonrelativistic regime, the Larmor radius of electrons remains small 
compared to the size of the scattering centers. The scattering center frame 
$\Rw$ and  the electron bulk velocity are, therefore, drifting at similar 
speeds (Fig.\ref{fig:fig1}g). In this diffusive limit,  the distribution 
function can be expanded in Legendre polynomials $f_s(p,\mu)\,=\,f^{(0)}_s(p)\,
+\,\mu f^{(1)}_s(p)$. Averaging Eq.~\eqref{eq:tr_eq} over the first two Legendre polynomials and 
assuming a dominant contribution from the electric field, 
we obtain a kinetic closure 
$f^{(1)}_{\rm e}\,\simeq\, -\frac{1}{\nu_{\rm e}}\,\left[ q_{\rm e} E_x  \,\partial_p f^{(0)} +\, \frac{p}{\me}\,\partial_x f^{(0)}\right]$ 
for the Fokker-Planck equation accounting for momentum diffusion in the electrostatic fields~\cite{Supp_Mat}:
\begin{align}
u_{\rm w} \partial_x f^{(0)} - \frac{p}{3}\,\partial_x u_{\rm w}\,\partial_p f^{(0)}=\frac{1}{3 p^2} \partial_p p^2 \frac{e^2 E_x^2}{\nu} \partial_p f^{(0)} .
\end{align}
The right-hand side only accounts for the dominant term responsible for the bulk heating 
of the electrons. This term predicts heating proportional to the diffusion coefficient 
$D_{pp} \sim \frac{1}{3}  e^2 E^2_x/\nu_{\rm e}$~\footnote{An analogous form was obtained in~\cite{Krimskii_1977} 
for the noninertial contribution uniquely and in ~\cite{Vanthieghem_2022, Vanthieghem_2022b} 
in the respective relativistic regimes of collisionless and radiation-mediated blast waves.  
}. 

Based on this picture, the development of a coherent electric field across the shock transition 
leads to electron stochastic heating. However, the origin of this electric field is not explicit. 
With a rate proportional to the square of the field amplitude, the heating mechanism we describe 
is similar to Joule heating from ambipolar diffusion. 
The decelerating scattering centers effectively act as a neutral species on which electrons and 
ions elastically scatter at different rates. An ambipolar electric field ensues from the larger 
effective frictional drag of the decelerating microturbulence on the electrons relative to the 
ions, leading to efficient diffusive heating of the electrons in a Joule process.

To capture the ambipolar nature of electron heating and the self-consistent coupling 
between electrons and ions, we combine a Monte Carlo (MC) method with a 
Poisson (MC-P) solver to achieve a complete solution of the transport equation~\cite{Vanthieghem_2019}. 
We use a resolution $\Delta x \,=\, 0.1\,c/\we$, $c\Delta t\,=\,0.99\,\Delta x$, $2^{\rm nd}$-order 
spline interpolation between particles and $E_x$-field, with isotropic white noise statistics 
for pitch-angle scattering. Electrons and ions are injected from the right-hand 
side of the domain with an initial bulk velocity $u_\infty\,=\,-0.075\,c$ and a temperature matching 
the initial conditions of PIC simulations.
For a fair comparison with PIC simulation, we used a 2D scattering operator.
Comparison between 2D and generalized 3D scattering operators showed no significant 
differences up to realistic mass ratios. Using our solver, we recover the Rankine-Hugoniot jump conditions 
in the corresponding dimension.
In the MC-P solutions, the scattering frequency is assumed constant, with values consistent with the analytical estimates discussed below. While the scattering frequency 
would have a spatial dependence due to the evolution of the microturbulence, the heating 
is dominated by regions of strong deceleration of $\Rw$ over the shock transition of size $L_{\rm sh}$.

We naturally expect two different scattering regimes to emerge for electrons and ions depending on the magnetization of the respective species. For the typical observed magnetic field strength, 
$|eB|\,\sim\,0.06\,(\tfrac{\mi}{\me})^{\scriptscriptstyle \frac{1}{2}}\,\me\we |\tfrac{u_\infty}{c}|$, and scale $k_\perp\,\sim\,(\tfrac{\mi}{\me})^{\scriptscriptstyle -\frac{1}{2}}\,\we/c$, produced by 
the Weibel instability (Fig.~\ref{fig:fig1}a), electrons moving at $u_\infty$ have a 
gyroradius much smaller than the size of the magnetic structures and are thus trapped. 
An estimate of their scattering frequency derives from the coherence time of the bounce frequency 
$\omega_{\beta}$ in the filaments of radius $r_\perp\,=\,2\pi/k_\perp$ and length 
$r_\parallel\,=\,2\pi/k_\parallel$~\cite{Lemoine_2019}. 
Explicitly, $\nu_{\rm e}\,\sim\,\Delta \alpha_{\rm e}^2/\Delta t_{\rm e}$ depends on the angle of deflection 
squared  $\Delta \alpha_{\rm e}^2 \,=\,\omega_{\rm \beta, e}^2\,r_\perp^2/u_{\rm th, e}^2$, 
where $u_{\rm th, e}$ is the thermal velocity, $\omega_{\beta, \rm e}\,\sim\,u_{\rm th, e}\sqrt{\,k_\perp/r_{\rm g,e} }$
is the bounce frequency of an electron~\cite{Ruyer_2018}, and on the coherence 
time $\Delta t_{\rm e}\,=\, 2 \pi/\left(k_\parallel u_{\rm th, e} \right)$.  
On the other hand, scattering of the high-rigidity ions propagating in the turbulence can be well approximated as a non-resonant process associated with small pitch-angle scattering with the usual estimate 
$\Delta \alpha^2_{\rm i}\,\sim\,r_\perp^2/r_{\rm g, i}^2$, with $r_{\rm g,i}$ the ion Larmor radius, and the shortest scattering time in the Weibel turbulence
$\Delta t_{\rm i}\,=\,2 \pi/\left\{\max\left[k_\parallel,k_\perp\right]|u_{\rm \infty}|\right\}$~\cite{Supp_Mat}. The interaction between the beam and incoming ions drives the turbulence, but the slowdown of the incoming ions defines the relevant scattering length. Close to the shock, the velocity $u_{\rm i}\,\sim\,u_\infty$, and thus $r_{\rm g,i}\,\sim\,\mi |u_\infty/eB|$. The analytical estimates for the scattering frequencies are then:
\begin{align}
    &\nue\,\simeq\,2 \pi\,\frac{k_\parallel}{k_\perp}\,\frac{\mi}{\me}\,\frac{|u_\infty|}{r_{\rm g,i}} \,, \label{eq:nue}\\
    &\nui\,\simeq\,\,\frac{r_\perp |u_\infty|}{r_{\rm g,i}^2}\,\max\left[ \frac{k_\parallel}{k_\perp},\,1 \right] \label{eq:nui}\,.
\end{align}

\begin{figure}
  \centering
  \includegraphics[width=1\columnwidth]{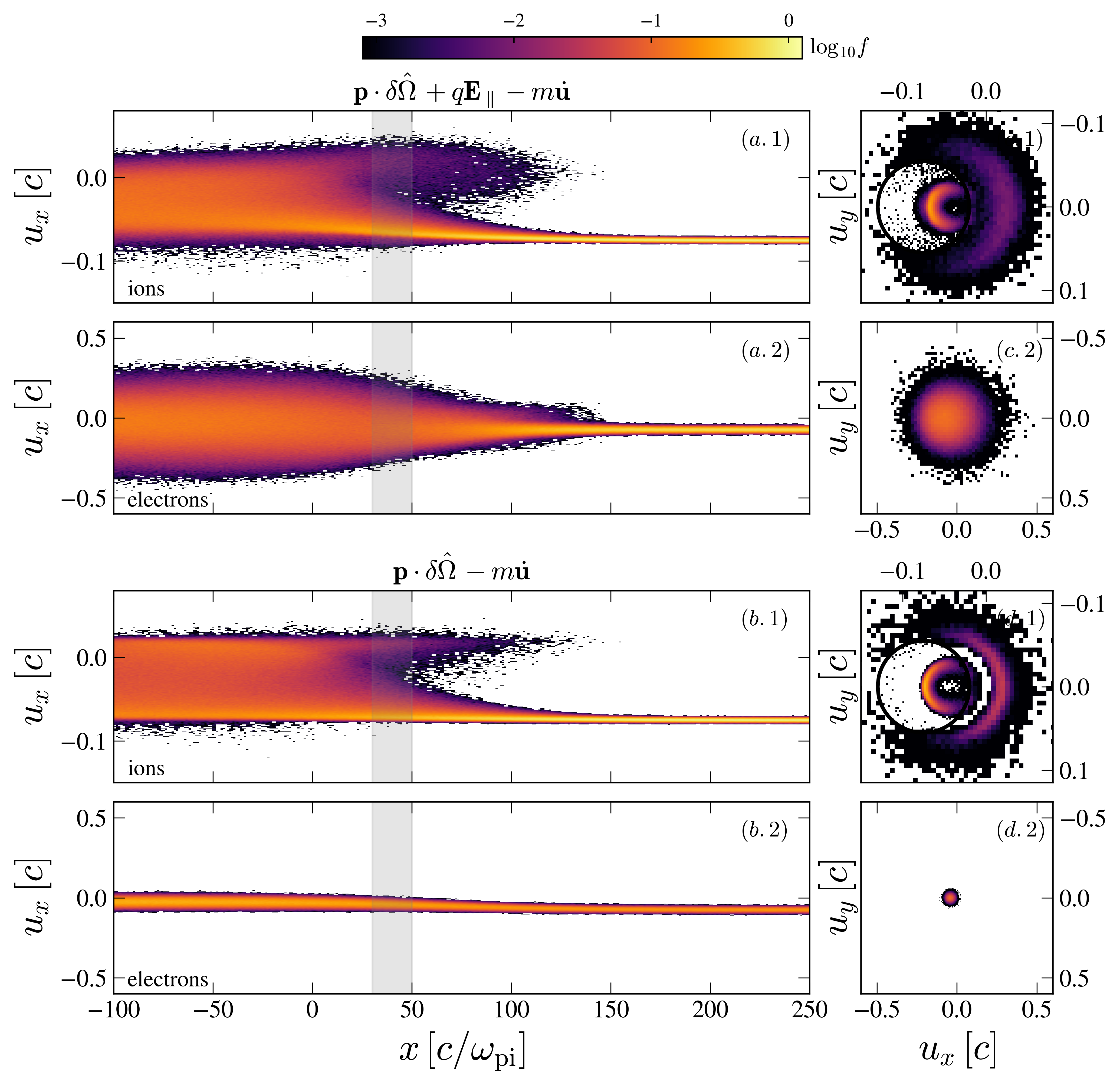}
  \caption{MC-P solution to the transport equation with $\nui\,=\,3.5\times10^{-4} \wi$ 
  and $\nue\,=\,8.5\times 10^{-3} \wi$ constant along the shock transition. The 
  upstream 4-velocity $u_\infty\,=\,-0.075$. The initial temperature matches the 
  PIC simulation. The top two panels correspond to the full MC-P solution for the 
  ions (a.1) and electrons (a.2). Panels (b.1-2) respectively show the equivalent 
  MC solution in the absence of longitudinal electrostatic field. For reference 
  and comparison to insets (d) and (e) of Fig.\ref{fig:fig1}, the insets on the 
  right show the corresponding phase space distribution in the shaded area. The 
  black circle marks the boundary between the definition of the beam and the 
  background plasma.} 
  \label{fig:fig2}
\end{figure}

In Fig.~\ref{fig:fig2}, we show the $x-p_x$ phase-space distributions for $\nue\,\simeq\,\nui \mi/\me$ 
obtained either from the full MC-P or MC solutions. 
Essentially, the second case reduces to the contribution of the purely motional electric field. 
In this case, electrons heat up almost adiabatically, trapped and compressed by the turbulence, 
ending with a negligible downstream temperature: $T_e/T_i \lesssim m_e/m_i$. When the full solution 
is considered, we observe that electrons are predominantly heated by the longitudinal electrostatic 
field, giving rise to a downstream temperature ratio $T_e/T_i \sim 0.5$, consistent with the full PIC simulations.
We note that the transport equation assumes small pitch-angle scattering for particles in 
microturbulence, which is valid for ions but not necessarily for electrons due to their smaller 
Larmor radius. We have checked that accounting for correlated scattering for electron transport does not significantly affect the dynamics~\cite{Supp_Mat}.

The transport equation~\eqref{eq:tr_eq} captures the essential electron dynamics. 
With the general goal of building a reduced model for the shock profile and electron 
heating, we now derive the set of fluid equations for the electron distribution.  
We note that the following equations and the previously introduced Fokker-Planck 
description are not supposed to be valid for ion species for which the diffusive 
approximation would fail across the shock transition. However, a closed form of the 
fluid equation for the electrons can be derived in the diffusive approximation from the 
first moments of the distribution.
We decompose the total stress-energy tensor $T_{(0)}^{xx}\,=\,\tau^{xx}_{(0)}\,+\,m u_{\rm w} j_{(0)}^x$
in terms of thermal pressure $\tau^{xx}_{(0)}$ and number current $j_{(0)}^x$ components. 
In the thermal part of the distribution, we assume that the scattering frequency is independent of the particle momentum~\cite{Krimskii_1981}. 
A more complex polynomial dependence of $\nue(p)$ would be relevant to electron injection and acceleration, %result in the contribution of 
%higher-order moments --- i.e., a nonthermal component --- 
which is left for future work. Neglecting anomalous heat transport 
$\frac{1}{\nu m} \partial_x^2 \tau_0^{xx}$, relevant to the high-energy component 
of the distribution, we obtain a solution for the conserved current across the 
shock transition $j_{(0)}^x / j_\infty^x\,=\,  1/(1 \,-\, \frac{e E_x}{\nu_{\rm e} m_{\rm e} u_{\rm w}} )$. 
Moments of Eq.~\eqref{eq:tr_eq} then give: 
\begin{equation}\label{eq:H_rate}
   \frac{1}{\mi}\partial_x  \tau_{\rm e (0)}^{xx} \,\simeq\, \frac{u_\infty}{u_{\rm w}} \frac{{j_{e(0)}^{x}}^2}{\phi_\infty}\frac{e^2 E_x^2}{\nu_{\rm e} m_{\rm e} u_{\rm w}} \,,
\end{equation}
where $\phi_\infty\,=\,\mi j_\infty u_\infty$ is the ram pressure at $+\infty$. With 
an explicit form for the dominant non-adiabatic heating rate of the electrons in terms of the amplitude of the coherent electrostatic field, 
we now derive a closed form for the electron-ion coupling.

The dynamics of electrons, trapped in the turbulence, is well characterized by the
diffusive approximation. We distinguish two extreme regimes for the ions: 
\emph{diffusive} ($\nu_{\rm i} \gg \partial_x u_{\rm w}$) and \emph{unscattered} regimes ($\nu_{\rm i}\,\sim\,0$). 
\emph{Unscattered} ions are only subject to the electric field deceleration 
and fully isotropize in the downstream. As verified in  PIC simulations, 
quasi-neutrality between the reflected beam charge density $\rho_{\rm b}$ and background 
contribution is observed in the full shock precursor.
Given a charge density 
profile $\rho_{\rm b}$ for the reflected beam and background ion velocity $u_{\rm i}$, we obtain 
$e E_x \,\simeq\, \nue \me \left[ u_{\rm w} - u_{\rm i} \, \frac{1 }{1 + \rho_{\rm b} u_{\rm i}/\left(e j_\infty \right) } \right]$~\cite{Supp_Mat}. 
The electric field then only results from the relative drift between electrons and ions if  $\rho_{\rm b}\,\ll\,e n_\infty$.  

Assuming a weak deceleration of ions across the shock transition size $L_{\rm sh}$ 
and neglecting the beam contribution, the system is fully parametrized by a single parameter 
$\xi\,=\,L_{\rm sh} \me \nue/\mi|u_\infty|$ that represents the number of electron 
scattering events across the shock transition times the electron-to-ion mass ratio. 
In the absence of another relevant scale, $\xi$ should be solely determined by the 
structure of the turbulent field. In such conditions, the scattering frequency of 
the ions determines the typical shock transition size $L_{\rm sh}\,\sim\,|u_\infty|/\nui$. 
Therefore, $\xi$ corresponds to the typical ratio between the electron and ion scattering frequencies, 
i.e., $\xi\,\sim\,\me\nue/\left(\mi\nui\right)$. 

For a linear deceleration profile of $\Rw$ in $x\,\in\,\left[- L_{\rm sh},0\right]$, 
we obtain $e E_x / \left( \me \nue u_\infty \right)\,\simeq\, \frac{3}{4\xi}\,\left( e^{\xi x /L_{\rm sh}} - 1 \right)$~\cite{Supp_Mat}. 
One can then derive the heating rate to the leading order in $\xi$:
\begin{align}
     \left|\partial_{x/L_{\rm sh}}  \tau_{e(0)}^{xx} \right| \,\simeq\, \phi_{\infty} \left\{ 
  \begin{array}{ c l }
    \frac{3}{16}\xi & \quad \textrm{if } \xi \lesssim 1\,, \\
    \, &\, \\
   \quad \frac{3}{4}\,\xi^{-1}    & \quad \textrm{if } \xi \gg 1\,,
  \end{array}
\right.
\end{align}
giving a fair estimate of the average heating across the shock transition when compared with the full integration.
\begin{figure}
  \centering
  \includegraphics[width=1\columnwidth]{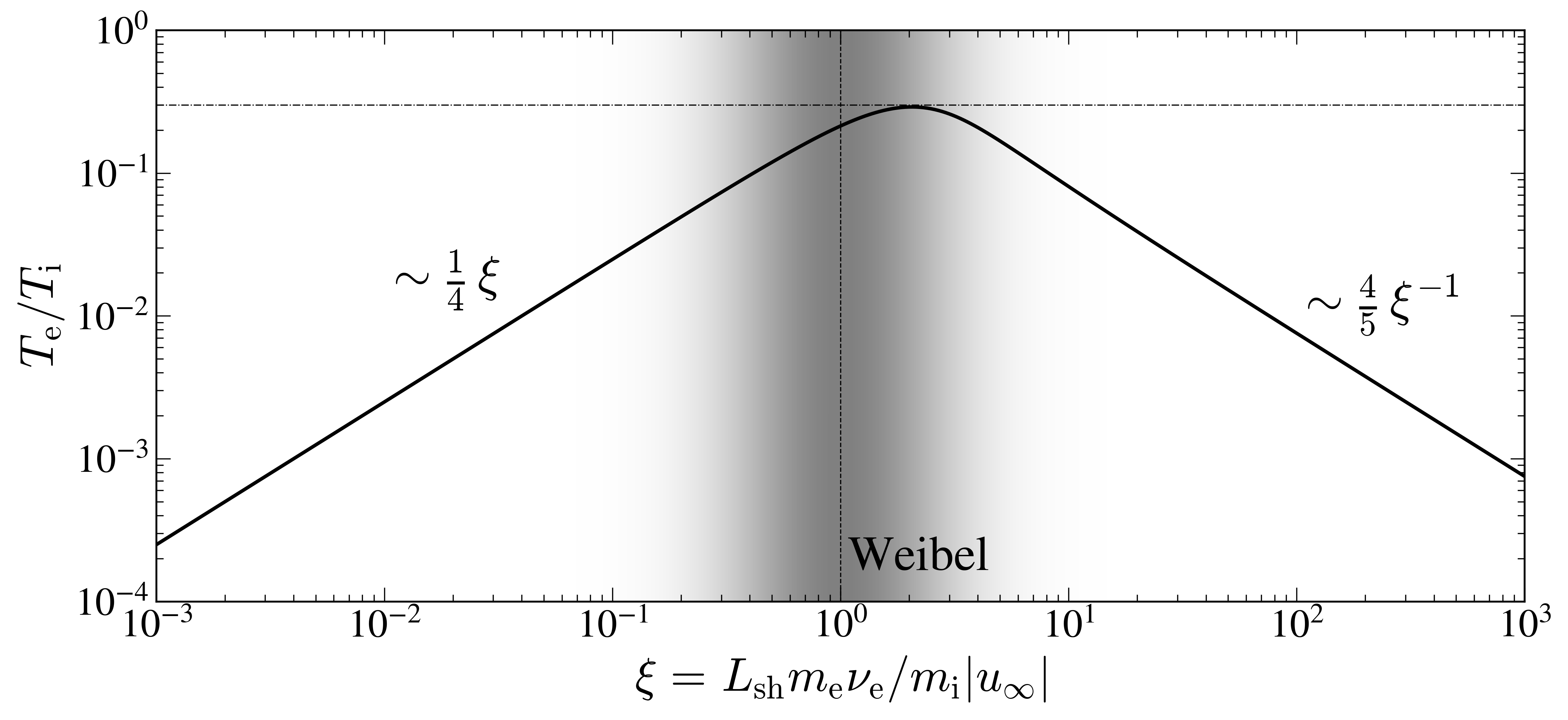}
  \caption{Estimate of the downstream temperature ratio for diffusive electrons and weakly 
  scattered ions over the shock transition in terms of the single free coupling parameter. 
  The temperature ratio is estimated using the downstream electron temperature given by 
  Eq.~\eqref{eq:H_rate} with the electric field profile derived for a linear deceleration. 
  The jump conditions constrain the ion pressure, assuming an isotropic distribution, 
  $\tau^{xx}_{\rm e} + \tau^{xx}_{\rm i} \,=\, \frac{3}{4} \phi_\infty $.}
  \label{fig:fig3}
\end{figure}
Provided that $\xi\,\lesssim\,1$, the downstream electron pressure is therefore of the order 
of $\tau_{e(0)}^{xx}\,\sim\, 0.2\,\xi\,\phi_\infty$. The full solution for a linear 
deceleration, depicted in Fig.~\ref{fig:fig3}, confirms that $\xi\,\sim\,1$ is necessary 
to recover downstream temperatures on the order of a fraction of unity. 

Interestingly, the scaling $\xi\,\sim\,\me\nue/\mi\nui$ is also observed 
in the fully diffusive regime for 
electrons and ions, suggesting generality of this scaling law.  Equations~\eqref{eq:nue} and~\eqref{eq:nui} 
provide direct estimates of the electron and ion scattering frequencies for the Weibel instability.  The ambipolar heating parameter then becomes
$\xi \,\sim \,  k_\perp\,r_{\rm g,i}\,\min\left[ \frac{k_\parallel}{k_\perp},\,1 \right]$.  
Using the values for filament scale in the shock transition, $k_\perp \sim k_\parallel \sim \wi/c$~\cite{Kato_2008, Ruyer_2018, Swadling_2020}, and the field amplitude set by trapping, $\omega_{\beta\rm, i}\,\sim\,|u_\infty|\,\sqrt{k_\perp/r_{\rm g,i}}\,\sim\,|u_\infty/c|\,\wi$~\cite{Davidson_1972}, we obtain $\xi\,\sim\,1$, precisely in the range of parameters maximizing the downstream electron temperature in Fig.~\ref{fig:fig3}. 
We note that the model developed here is general in that it could 
also be applied to other nonlinear processes changing the properties of the 
magnetic turbulence upstream of the shock, such as non-resonant current-driven instabilities~\cite{Bell_2004},
merging~\cite{Medvedev_2005,Vanthieghem_2018,Zhou_2020}, cavitation~\cite{Peterson_2021}, reconnection~\cite{Matsumoto_2017,Bohdan_2020}, or kink modes~\cite{Ruyer_2018}. This would impact $T_{\rm e}/T_{\rm i}$ primarily through the scale of the 
turbulent magnetic field -- i.e., $k_\parallel$ and $k_\perp$. For example, if the field saturates at the scale of the dominant Larmor radius of the beam $k\,r_{\rm g, b}\,\sim\,1$, untrapped ions result in $\xi\,\sim\,\frac{r_{\rm g,i}}{r_{\rm g,b}}$, and the temperature ratio decreases with $r_{\rm g,b}$. However, if these late nonlinear modes ultimately lead to efficient ion trapping, then we naturally recover $\xi\,\sim\,1$, such that the temperature ratio becomes weakly sensitive to subsequent nonlinearities.

In summary, we have developed a self-consistent model for the energy partition 
in high Mach number collisionless blast waves. The heating results from the
ambipolar electric field that accelerates electrons, which are thermalized by rapid scattering 
in the Weibel-mediated turbulence. We find that the downstream temperature ratio can be expressed in terms of a single dimensionless parameter determined by the nature of the 
dominant instability. The heating rate and temperature ratio between electrons 
and ions exhibit good agreement with ab initio fully kinetic simulations, 
semi-analytical MC-P solutions, and reduced analytical models. Energy partition 
peaks around $T_{\rm e}/T_{\rm i}\,\sim\,0.3$ with a weak dependence on higher-order 
effects in the statistics of electron transport and nonlinear dynamics of the instability.  
Our model gives a natural interpretation for the thermal partition in shocks particularly 
relevant to weakly magnetized astrophysical systems and to ongoing laboratory 
experimental studies. More generally, these findings also open promising avenues 
for studying electron transport in magnetically dominated systems, for which 
$L_{\rm sh}$ is set by the coherent Larmor gyroradius, and potential electron 
injection and acceleration in turbulent shocks. 

%%%%%%%%%%%%%%%%%%%%%
% Acknowledgments
%%%%%%%%%%%%%%%%%%%%%
\begin{acknowledgments}
\emph{Acknowledgments} --- 
AV acknowledges M. Medvedev, A. Bohdan, S. Okamura, M. Toida, and L. Sironi for insightful discussions. 
AV acknowledges H. Nagataki and H. Ito at RIKEN and T. Amano and Y. Ohira at the University of Tokyo 
for their hospitality during the conclusion of this work. 
AV acknowledges support from the NSF grant AST-1814708 and the NIFS Collaboration 
Research Program (NIFS22KIST020). FF acknowledges support from the European Research Council 
(ERC-2021-CoG Grant XPACE No. 101045172). This research was facilitated by the Multimessenger 
Plasma Physics Center (MPPC), NSF grant PHY-2206607. The presented numerical simulations were 
conducted on the stellar cluster (Princeton Research Computing). This work was granted access to 
the HPC resources of TGCC/CCRT under the allocation 2024-AD010415130R1 made by GENCI.
\end{acknowledgments}

\bibliography{bib_letter.bib}

\newpage

\begin{center}
 --- \Large\textbf{Supplementary Information}  ---
\end{center}

%%%%%%%%%%%%%%%%%%%%%
% Content
%%%%%%%%%%%%%%%%%%%%%
\section{Transport in the diffusive limit} \label{sec:red_noise}

We detail the electron semi-dynamical equation and closure for electrons in the 
diffusive limit. The interaction with the Weibel microturbulence first reduces 
to a pure pitch angle scattering in the scattering center frame $\mathcal{R}_{\rm w}$. 
The equation of motion for a single particle reads:
\begin{align} \label{eq:SDE}
  \dot{\bm{p}} \,=\, \bm{p} \cdot \, \delta\hat{\Omega}_t + q\, \bm{E} - m \dot{\bm{u}}_{\rm w} \,,
\end{align}
where $u_{\rm w}$ is the drift velocity of the $\mathcal{R}_{\rm w}$ in the shock front
frame and $\delta\hat{\Omega}_t$ is a stochastic process. The stochastic variable 
$\delta\hat{\Omega}_t$ is a linear combination of the generators of the Lie algebra of 
the rotation group corresponding to pitch-angle scattering in the turbulent magnetic 
field~\cite{Plotnikov_2011}. In other words, the infinitesimal scattering operator is 
decomposed on the generators of the Lie algebra:
\begin{align}
    \varpi_1\,&=\, \begin{pmatrix}
0 & -1 & 0\\
1 & 0 & 0\\
0 & 0 & 0
\end{pmatrix} \,,\quad\varpi_2\,=\, \begin{pmatrix}
0 & 0 & 1\\
0 & 0 & 0\\
-1 & 0 & 0
\end{pmatrix}\,,\quad\\
&\qquad\qquad\varpi_3\,=\, \begin{pmatrix}
0 & 0 & 0\\
0 & 0 & -1\\
0 & 1 & 0
\end{pmatrix}\,.
\end{align}
We can now compute the diffusion coefficient of the transport equation deriving from 
the above Langevin equations in the respective dimensions. Note that drift and diffusion 
coefficients are uniquely determined from the Langevin equation. To do so, it is convenient 
to perform a change in variables $(p^x,p^y,p^z)\,\to\,(p,\mu=\cos\theta,\phi)$ where $\mu$ 
is the pitch-angle cosine. The SDE for momentum then reduces to
\begin{equation}
\dot{p}\,=\,  \mu \, \left(  q E^x - m \dot{u}_{\rm w} \right)\,,
\end{equation}and the random part of the pitch-angle variation respectively reduces to
\begin{align}
&\dot{\theta}_{\rm st}\,=\, \chi_t\,,   \qquad \text{(2D)}\,\\
&\dot{\mu}_{\rm st}\,=\, \frac{p^y}{p} \chi^y_t + \frac{p^z}{p} \chi^z_t\,,  \qquad \text{(3D)}\,
\end{align}
such that, for isotropic pitch-angle scattering, the diffusion coefficients reduce 
to $D_{\theta\theta}^{(2D)}\,=\,\nu$ and $D_{\mu\mu}^{(3D)}\,=\,\nu\,\left( 1 - \mu^2\right)$~\cite{Risken}. 
The auto-correlation function 
\begin{align}
  \phi(t,t') \,=\, \langle \delta\hat{\Omega}_t \delta\hat{\Omega}_{t'} \rangle  \,,
\end{align}
depends on the nature of the stochastic process. In the standard case of a Gaussian 
white noise, the correlation function reduces to 
\begin{align}
  \phi(t,t') \,=\, 2\,\nu\,\delta(t'-t)\,, \label{eq:autocorr_white}
\end{align}
where $\nu$ is the pitch-angle scattering frequency. 
The differential equation~\eqref{eq:SDE} is equivalent to the following transport equation in
the shock-front frame where the system is assumed stationary:
\begin{align}\label{eq:tr_eq}
   \left( m \beta_{\rm w}  + p^x \right) \partial_x f_s &-\,m \partial_x u_{\rm w} \left( u_{\rm w} 
   m + p^x \right)  \partial_{p^x} f_s \nonumber \\
   +\,m\,q E_x\,\partial_{p^x} f_s  
   &\,=\, \frac{m}{2} \partial_\mu \left[ \nu_s \left( 1 - \mu^2 \right) \right] \partial_\mu f_s \,,
\end{align}
where the right-hand side is the operator for elastic scattering in $\mathcal{R}_{\rm w}$
in 3D with $\mu\,=\,\cos\theta$ the pitch-angle cosine. For a two-dimensional distribution, the 
operator reduces to $\partial_\theta\nu\partial_\theta f_s$. In the diffusive limit, the 
distribution function is expanded in Legendre polynomials $f_s(p,\mu)\,=\,f^{(0)}_s(p)\,
+\,\mu f^{(1)}_s(p)$. The transport of electrons is then modeled as a Fokker-Planck process. 
The above equation, projected along the first Lagrange polynomial, gives:
\begin{align}
  &u_{\rm w} \fo - \frac{1}{3} p \partial_x u_{\rm w} \partial_p \fo  \,=\, - \frac{p}{3 m} \partial_x \fl \nonumber \\
  & \qquad-\frac{1}{3} \left( \frac{2}{p} \fl + \partial_p \fl \right) \left(q E_x - \frac{1}{2} \partial_x u_{\rm w}^2 \right) \,. \label{eq:tr_eq_0}\\
  &\frac{p}{3 m} \partial_x \fo + \frac{1}{3} \left(q E_x - \frac{1}{2} \partial_x u_{\rm w}^2 \right) \partial_p \fo \,=\, - \frac{u_{\rm w}}{3}  \partial_x \fl\nonumber \\
  & \qquad- \frac{1}{3} \left( \nu + \frac{2}{5} \partial_x u_{\rm w} \right) \fl  + \frac{1}{5} p \partial_x u_{\rm w} \partial_p \fl  \,.
\end{align} 
Neglecting terms in $\partial_p f^{(1)}$ gives the following approximations for $f^{(1)}$:
\begin{align} \label{eq:f1}
    &f^{(1)}\,=\, -\frac{1}{\nu}\,\left[ \left( q E_x - \frac{m}{2} \partial_x u_{\rm w}^2 \right) \,\partial_p f^{(0)} +\, \right.\nonumber \\ 
&\quad\left. \frac{p}{m}\,\partial_x f^{(0)}\right] \,+\, \mathcal{O}\left[\frac{\left (\partial_x u_{\rm w} \right)^2}{\nu}\right]\,
\end{align}
which can be inserted in~\eqref{eq:tr_eq_0} to obtain a Fokker-Planck equation. We only consider 
the leading terms of the equation, assuming that the contribution from the electric field is 
dominant. The right-hand side components correspond to diffusive heating in the coherent 
electrostatic field, spatial diffusion, and cross-momentum-space diffusion. 
Core heating of the electrons is mainly affected by the first term, while 
the second and third components affect the high-energy part of the distribution. 

\section{Scattering and correlation} \label{sec:red_noise}

\begin{figure}
  \centering
  \includegraphics[width=0.95\columnwidth]{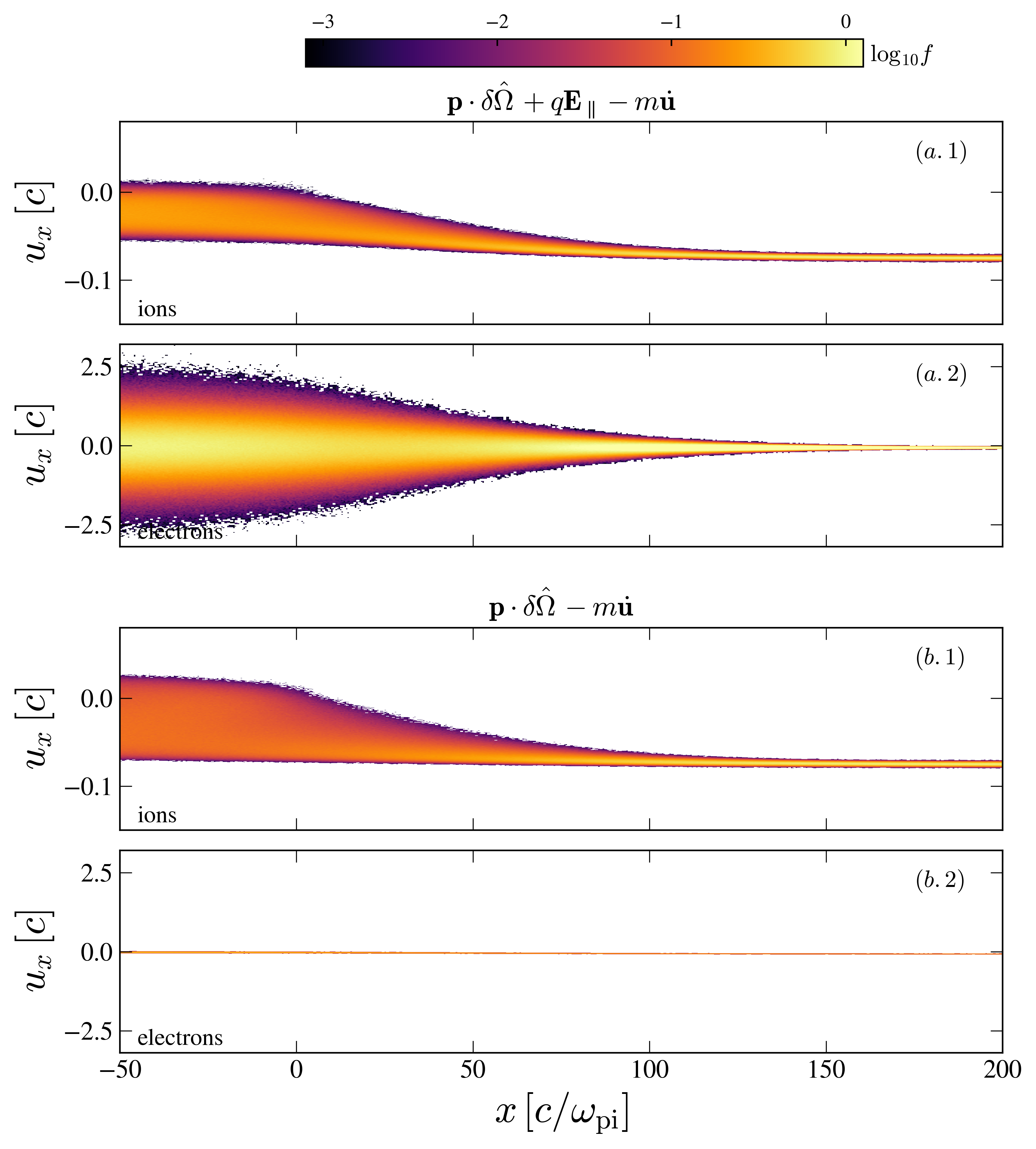}
  \caption{Same as Fig.~(2) in the letter but for a realistic mass ratio $\mi\,=\,1836\,\me$ and $\nue\,=\,1836\,\,\nui$. MC-P solution to the transport equation with $\nui\,=\,3.5\times10^{-4} \wi$, constant along the shock transition. The upstream 4-velocity $u_\infty\,=\,-0.075$. The top two panels correspond to the full MC-P solution for the ions (a.1) and electrons (a.2). Panels (b.1-2) respectively show the equivalent MC solution in the absence of a longitudinal electrostatic field. The temperature ratio in the downstream is $T_{\rm e}/T_{\rm i}\,\simeq\,0.5$ for the case (a).}
  \label{fig:2_SM}
\end{figure}

\begin{figure}
  \centering
  \includegraphics[width=1.\columnwidth]{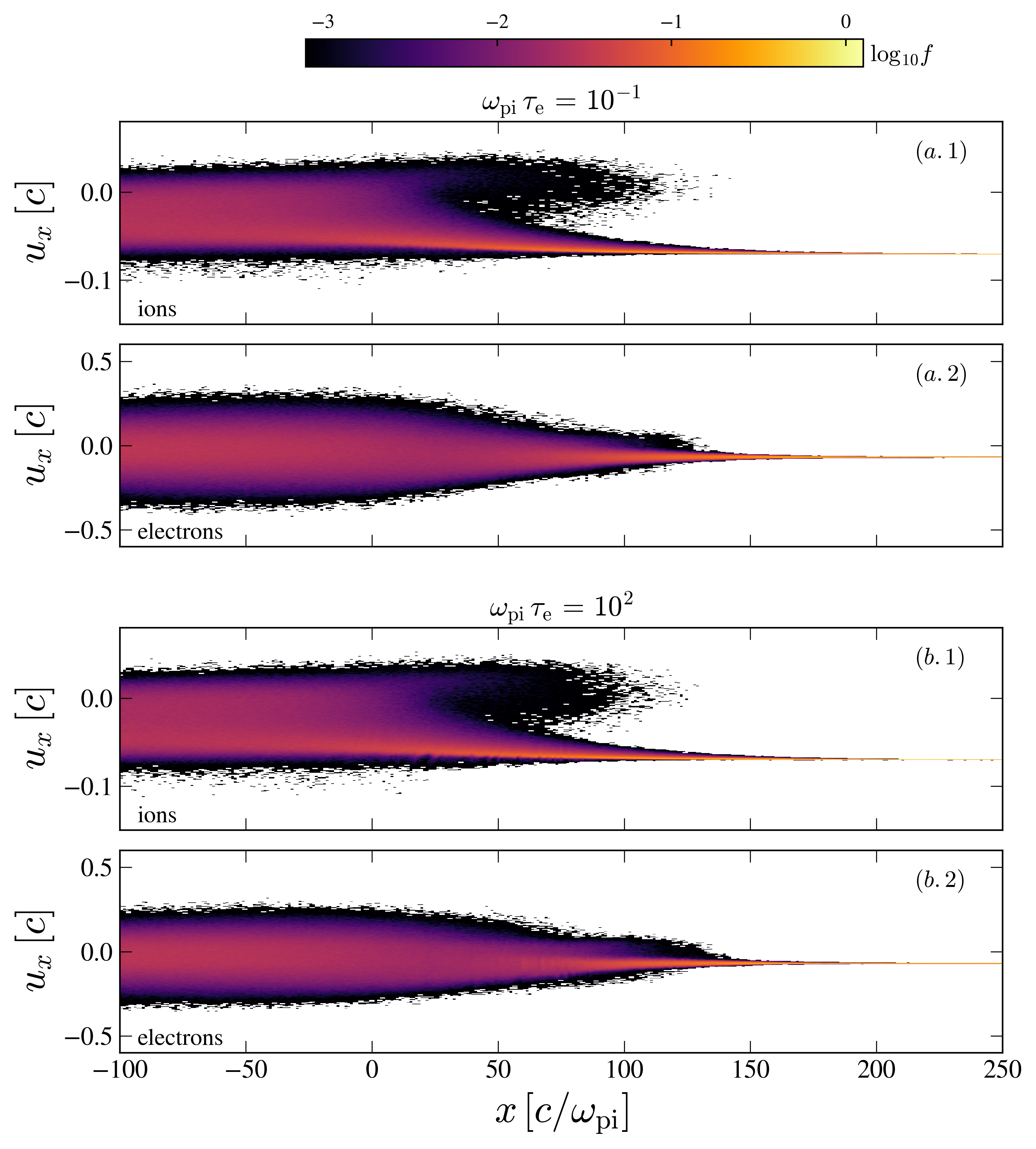}
  \caption{MC-P solution to the transport equation with $\nui\,=\,3.5\times10^{-4} \wi$ and 
  $\nue\,=\,8.5\times 10^{-3} \wi$ constant along the shock transition for the two respective 
  correlation times $\wi\,\tau\,=\,10^{-1}$ [see (a.1-2)] and $\wi\,\tau\,=\,10^2$ [see (b.1-2)] 
  for a red-noise type scattering in pitch-angle. The odd and even panels, respectively, 
  correspond to the ion and electron phase spaces. The shock transition is centered on $\wi x/c\,=\,0$.}
  \label{fig:MCP_corr}
\end{figure}

Integration of the Langevin equation using the MC-P solver demonstrates efficient electron heating through the ambipolar electric field. The main free parameter is the scattering frequency of the particles, estimated analytically in both the trapped and untrapped regimes. To motivate their value, we extract $\nu_{\rm i}$ from PIC simulations using two standard estimates $\langle\frac{1}{2} {\rm d}\theta^2/{\rm d} t \rangle$ and 
$\lim_{t \to \infty} \langle \theta^2(t) \rangle/2 t$, which implicitly neglects the spatial dependence of the scattering coefficient~\cite{Balescu_1997}. The pitch angle $\theta$ is measured, accounting for the phase of the particle as it gyrates inside the turbulence. Measurement of the scattering frequency over $2000$ tracked particles over $1.6\times10^5$ time-steps indicates $\nui\sim\,10^{-4}-10^{-5}\,\wi$ across the shock precursor and continuously increases toward the shock transition through the growth of the magnetic field in scale and amplitude. Consistent with the analytical estimates.

The efficiency of the heating mechanism up to a realistic mass ratio is demonstrated with a solution to the MC-P approach in Fig.~\ref{fig:2_SM}. We note that the temperature ratio of $T_{\rm e}/T_{\rm i}\,\sim\,0.5$ in the series of MC-P solution is slightly larger than the one obtained from PIC simulations where $T_{\rm e}/T_{\rm i}\,\sim\,0.3$. While the same electron temperature and dynamics are recovered with $k_{\rm B} T_{\rm e}\,\simeq\,0.1 \tfrac{1}{2}\mi u^2_\infty$, the ions are slightly colder in the MC-P solution such that the temperature ratio is larger.

The white noise limit associated with the transport equation~\eqref{eq:tr_eq} assumes particles undergoing uncorrelated small angle scattering in the microturbulence. This approximation is fair for ions. However, the electron Larmor radius is comparable to or smaller than the typical size of the filaments, and variations in pitch angle are not necessarily independent of time. We now investigate the effect of correlation of the pitch-angle scattering for particles of Larmor radius smaller than the typical scale of the turbulence~\cite{Deligny_2021}.
In such a case, the auto-correlation function deviates from 
~\eqref{eq:autocorr_white}. 
We hereafter assume the autocorrelation function $\phi(t,t')$ satisfying:
\begin{align}
  \phi(t,t') \,\propto\,\exp\left( -\frac{t' - t}{\tau} \right)\,,
\end{align}
where $\tau$ is the correlation time of the pitch-angle variation in the microturbulence.
The motivation behind the use of colored noise thus becomes more transparent: between scattering events, electrons undergo a series of coherent betatron oscillations and progressively lose coherence as they transition from one structure to another.

To test the effect of correlation, we scatter electrons, assuming an exponentially correlated scattering operator. A colored noise with zero mean and an exponential auto-correlation profile is called a red noise. We obtain the correlated noise from a white noise $\zeta_t$ using the straightforward relation:
\begin{align}
  \delta\dot{\hat{\Omega}}_t \,=\, - \frac{\delta \hat{\Omega}_t}{\tau} + \frac{\chi_t}{\tau}\,.
\end{align}
We integrate the transport equation with a finite correlation time for the electrons varying between $\wi\,\tau\,=\,10^{-1}$ 
to $\wi\,\tau\,=\,10^{2}$, see Fig.~\ref{fig:MCP_corr}. For $\wi\,\tau_{\rm e}\,=\,10^{-1}$, we recover the limit of white noise as expected for correlation time much shorter than the scattering time in the turbulence. The result of the integration is shown in Fig.~\ref{fig:MCP_corr}. The profile does not show a significant difference in electron heating over the shock transition. We thus conclude that the standard white noise limit captures the essential mechanism of electron heating.

\section{Fluid description and downstream temperature} \label{sec:fluid}

A closed form of the fluid equation can also be derived from the first moments of the distribution. 
The general form of the moment of order $N$, in the shock front frame, reads:
\begin{equation}
    \mathcal{M}_{| \rm sh}^{\alpha_1 \dots \alpha_N} \,=\, \int\,\frac{d^3p_{| \rm w}}{m} p^{\alpha_1}_{|\rm sh} \dots p^{\alpha_N}_{|\rm sh} \,f(p_{|\rm w},\mu_{|\rm w})\, 
\end{equation}
The current and the longitudinal momentum flux are then, respectively, 
given by:
\begin{align}
  j^x\,&=\, 4 \pi u_{\rm w} \int {\rm d}p\, p^2 f^{(0)} \,+\, \frac{4 \pi}{3} \int {\rm d}p\, \frac{p^3}{m} f^{(1)} \,, \\
  &=\, j^x_{0} + j^x_{1} \,,\\
  T^{x x}\,&=\, 4 \pi \int {\rm d}p\, p^2 m \left( u^2_{\rm w} + \frac{1}{3} \frac{p^2}{m^2 } \right) f^{(0)} \nonumber \\
  &\qquad\qquad\qquad\qquad\,+\, \frac{8 \pi }{3} u_{\rm w} \int {\rm d}p\, p^3 f^{(1)} \,,\\
  &=\, T^{x x}_{0} + T^{x x}_{1}\,,
\end{align}
from which we obtain the following relations
\begin{align}
  &\frac{4 \pi}{3} \int {\rm d}p\,p^4 f_0\,=\, m \left( T^{xx}_0  -  m u_{\rm w} j^x_0 \right)  \,, \\
  &T^{xx}_1 \,=\, 2 m u_{\rm w} j^x_1  \,.
\end{align}
From the moments of the transport equation~\eqref{eq:tr_eq}, we obtain the following:
\begin{align}
  &\partial_x j^x_0 \,=\, - \partial_x j^x_1 \,, \label{eq:cont} \\ 
  &\partial_x T^{x x}_0 - q E_x  \frac{j^x_0}{ u_{\rm w} }   \,=\, - 
  \partial_x T^{x x}_1 - \frac{\nu }{2} \frac{T^{x x}_1}{u_{\rm w}} \label{eq:mom}\,. 
\end{align}
To ease comparison, we decompose the longitudinal momentum flux into thermal and ram pressure components such that:
\begin{align}
  \tau^{xx}_0 \,=\, T_0^{xx} \,-\, m u_{\rm w} j_0^x \,.
\end{align}
From higher moments of the transport equation and neglecting the contribution 
of the heat flux, we obtain the following closure relation:
\begin{align} \label{eq:closure}
 \frac{\partial_x \tau^{xx}_0}{\tau^{xx}_0} + \frac{5}{3} \frac{\partial_x u_{\rm w}}{u_{\rm w}}  \,\simeq\, \frac{2}{3} \frac{j_1^x}{\tau^{xx}_0 u_{\rm w}} \left( q E_x -\frac{m}{2} \partial_x u_w^2 \right) \,.
\end{align}
The above Eq.~\eqref{eq:closure} clearly shows the nonadiabatic nature of the turbulent
heating in a Joule-like process from diffusion in the coherent electric potential and 
deceleration of the plasma microturbulence. A closed form of the equation can also be 
obtained from the kinetic distribution in Eq.~\eqref{eq:f1},  assuming $\nu$ to be 
independent of $p$. The first moment of Eq.~\eqref{eq:f1} readily gives:
\begin{align} \label{eq:j1}
    j_1^x \,\simeq\, \frac{1}{\nu m}\,\left[ \left(q E_x - \frac{m}{2} \partial_x u_{\rm w}^2\right) \frac{j_0^x}{ u_{\rm w}}  - \partial_x \tau_0^{xx}\right] \,.
\end{align}
Inserting this closure into Eq.~\eqref{eq:cont} to leading order in $|E_x|$, we obtain:
\begin{align} \label{eq:j0}
    \left( 1 + \frac{q E_x}{\nu m u_{\rm w} }\right) \partial j_0^x \,=\, - \partial_x \left( \frac{q E_x}{\nu m u_{\rm w} } \right) \, j_0^x + \frac{1}{\nu m} \, \partial_x^2 \tau_0^{xx} \,.
\end{align} 
Neglecting $\partial_x^2 \tau_0^{xx}$ in Eq.~\eqref{eq:j0}, we directly have a solution for 
$j_{e(0)}^x / j_\infty^x\,=\,  1/(1 \,-\, \frac{e E_x}{\nu_{\rm e} m_{\rm e} u_{\rm w}} )$. 
From quasi-neutrality $-e j^{x}_{\rm e(0)}/u_{\rm w}\,+\,e j^x_{\infty}/u_{\rm i} \,+\, \rho_b\,\simeq\,0 $,
we then obtain an estimate of the ambipolar field in terms of the Weibel frame velocity,
the ion velocity, and the charge density of the beam of reflected ions $\rho_{\rm b}$. The rest of the calculations leading to Eq. (5) in the letter are outlined in the main text. Using the relations we derived, we now discuss the direct estimate of the electric field across 
the shock transition. We consider a linear deceleration region between 
$x\,\in\,[-\frac{1}{2},\frac{1}{2}]\,L_{\rm sh}$ of the form:
\begin{align}
    u_{\rm w} \,=\, u_\infty\,\left(\frac{5}{8} + \frac{3}{4} \frac{x}{L_{\rm sh}} \right) \,.
\end{align}
With a linear deceleration profile and neglecting ion scattering and beam contribution 
across the shock transition, electron pressure is directly obtained from the following 
set of equations:
\begin{align}
    &\frac{1}{2} \partial_{x/L_{\rm sh}} \frac{u_i^2}{u_\infty^2}\,=\, \frac{u_\infty}{|u_\infty|} \left( \frac{u_{\rm w}}{u_\infty} - \frac{u_i}{u_\infty} \right)\,\xi \,, \label{eq:E_an}\\
    &\partial_{x/L_{\rm sh}} \tau^{xx}_{\rm e (0)} \,=\, \phi_\infty \frac{u_\infty}{|u_\infty|} \left( \frac{ \frac{|u_\infty|}{u_{\rm w}} \, \frac{1}{2} \partial_{x/L_{\rm sh}} \frac{u_i^2}{u_\infty^2} }{1 - \frac{|u_\infty|}{u_{\rm w}} \, \frac{1}{2} \partial_{x/L_{\rm sh}} \frac{u_i^2}{u_\infty^2}} \right)^2\,\xi \,. \label{eq:T_an}
\end{align}
The numerical solution, shown in the main paper, peaks at $T_{\rm e}/T_{\rm i}\,\sim\,0.3$ for $\xi\,\sim\,2$, in good agreement with particle-in-cell simulations. 

\begin{figure}
  \centering
  \includegraphics[width=0.95\columnwidth]{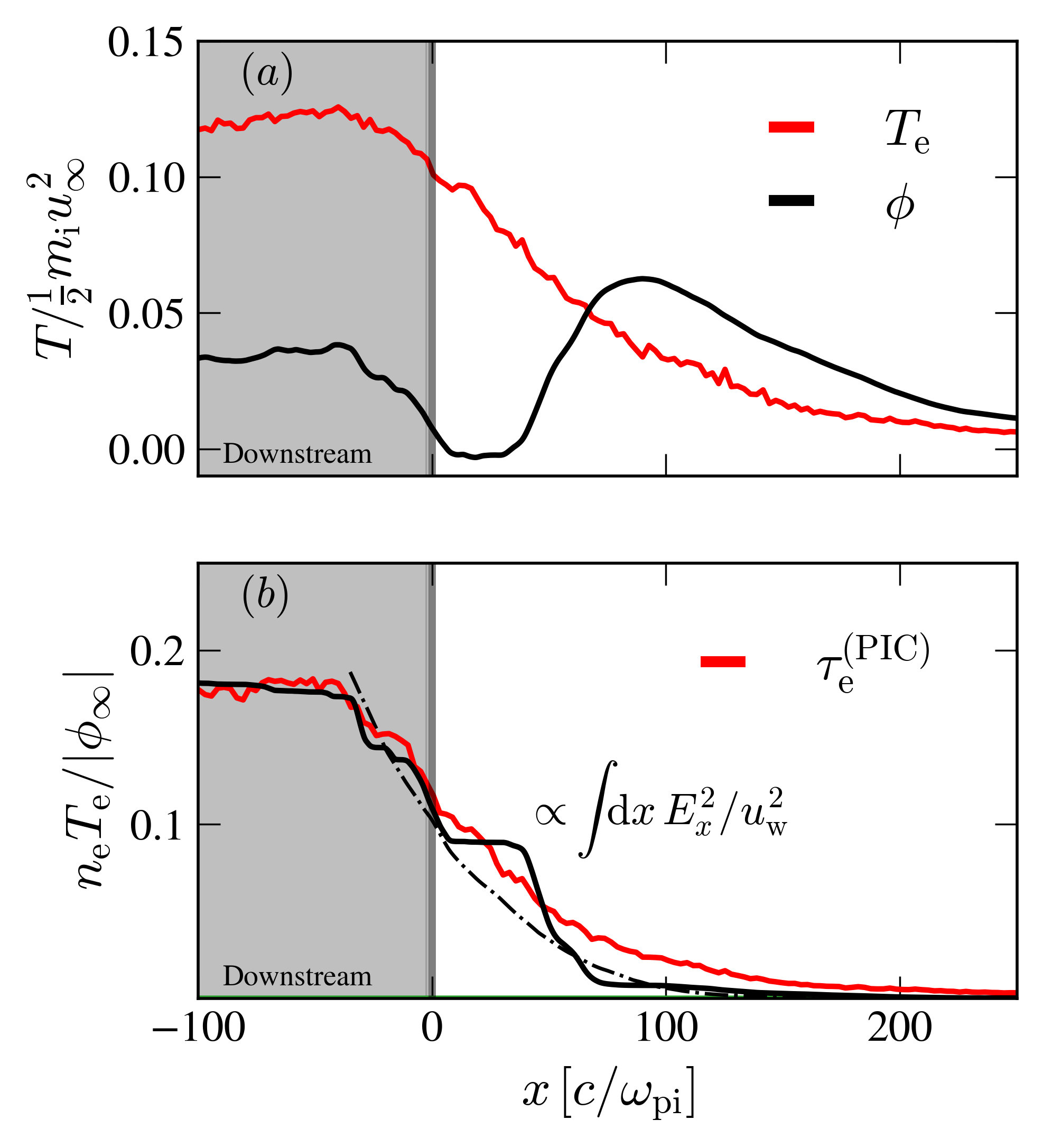}
  \caption{Top panel: fraction of the incoming kinetic energy conferred to the electron 
  population (red) and time-average electric potential (black) normalized by the incoming 
  kinetic energy of the ions. Bottom panel: Comparison between the electron kinetic 
  pressure measured in the simulation (red), the theoretical estimate obtained in 
  Eq. (3) up to a prefactor of order unity ($\sim 3/2$) to match the exact downstream temperature to account for the different degrees of freedom between the 2D and 3D scattering operators and the unknown exact electron scattering frequency (solid black), and from the analytical estimate in Eqs.~\eqref{eq:E_an}-\eqref{eq:T_an} (dashed black).  The shaded area corresponds to the shock downstream.}
  \label{fig:temp}
\end{figure}
The diffusion coefficient, proportional to $E_x^2$, leads to a heating rate independent of the sign of the electric field. This is observed in  Fig.~\ref{fig:temp}.a, where the change of sign in $E_x$, corresponding to the modulation of $\phi$, is attributed to the influence of the beam captured in the full expression $e E_x \,\simeq\, \nue \me \left[ u_{\rm w} - u_{\rm i} \, \frac{1 }{1 + \rho_{\rm b} u_{\rm i}/\left(e j_\infty \right) } \right]$. In a region where $\left|\,e j_\infty\left(u_{\rm w} - u_{\rm i}\right)\right|\,\lesssim\, \rho_{\rm b} u_{\rm i}^2$, the sign of the electric field will experience a reversal, as observed in the simulation. However, adiabatic cooling is negligible due to the diffusive nature of the heating. To illustrate this, a comparison between the predicted and measured electron pressure is shown in the bottom panel Fig.~\ref{fig:temp}. The predicted pressure is multiplied by an arbitrary factor of order unity accounting for corrections to the dimensionality of the scattering operator and amplitude of the scattering frequency. The profile of the electrostatic potential obtained from the PIC simulation is shown in the top panel. Our reduced analytical model neglects the contribution of the beam in the total heating rate. To validate this hypothesis, Fig.~\ref{fig:temp}.b shows the associated analytical temperature profile obtained by direct integration of Eqs.~\eqref{eq:E_an}-\eqref{eq:T_an}. The ion and Weibel velocity profiles were extracted directly from the PIC simulation to close the system. The analytical profile captures the trend and amplitude of energy partition between species.

\section{Convergence} \label{sec:PIC}

To support the analytical and semi-analytical models described in the letter, we performed a series of 2D3V simulations with the massively parallel kinetic code Tristan-MP~\cite{Spitkovsky_2005}. Here, we describe the series of tests ensuring convergence of the electron temperature in the shock downstream. With run A, the main simulation, and a constant $c\Delta t = 0.45 \Delta x$, we vary the numerical resolution within $\Delta x \,=\,[0.05, 0.1]\,c/\we$ in run B, increased the shock velocity from $v_{\rm sh}\,=\,0.075 c$ to $v_{\rm sh}\,=\,0.225 c$ in run B, and increased the phase-space resolution in run C. We found a good convergence of the electron temperature.

\begin{table}[htbp]
    \centering
    \label{tab:messier}
    \begin{tabular}{cccccccc}
        \hline
        Run & $\Delta x$ & $L_y$ & $v_{\mathrm{sh}}$ & $m_i$ & $\mathrm{N_{ppc}}$ & $k_B T_e$  \\
            & $[c/\we]$    & $[c/\we]$ & $[c]$ & $[\me]$ & & $[\frac{1}{2}m_i u_{\mathrm{sh}}^2]$ \\[0.2cm]
        \hline
        A & $0.10$ & $400$ & $0.075$ & $49$ & $32$ & $0.099$  \\
        B & $0.10$ & $200$ & $0.225$ & $49$ & $32$ & $0.113$  \\
        C & $0.10$ & $200$ & $0.075$ & $49$ & $128$ & $0.098$ \\
        D & $0.05$ & $100$ & $0.075$ & $49$ & $32$ & $0.097$  \\
        \hline
    \end{tabular}
    \caption{Bench-marking of PIC simulations at $\omega_{p,e}t = 1.25\times 10^4$. 
    Grid resolution $\Delta x$ and transverse grid size $L_y$ are measured in 
    $c/\omega_{p,e}$. In all cases, we observe a downstream electron temperature 
    $k_B T_e \simeq 0.1 \frac{1}{2}m_i u_{\mathrm{sh}}^2$.}
\end{table}

\end{document}